\documentclass[aps,pre,preprint,showpacs,floatfix]{revtex4}

\usepackage{amsmath}
\usepackage{amssymb}
\usepackage{graphicx}
\usepackage{bm}
\usepackage{mathrsfs}

\begin{document}

\title{Incoherent interaction of light with electron-acoustic waves}  

\author{Mattias Marklund and Padma K.\ Shukla}
\affiliation{Department of Physics, Ume{\aa} University, SE--901 87 Ume{\aa},
Sweden} 

\date{\today}

\begin{abstract}
The equations governing the interaction  between incoherent
light and electron-acoustic waves are presented. The modulational instability properties
of the system are studied, and the effect of partially coherent light is
discussed. It is shown that partial coherence of the light suppresses the modulational
instability. However, short wavelength perturbations are less affected, and will
therefore dominate in, e.g.\ pulse filamentation. The results may be of 
importance to space plasmas and laser-plasma systems.
\end{abstract}

\maketitle

The investigation of the interaction between light and ion-acoustic waves was 
pioneered by Karpman in the early seventies \cite{Karpman1,Karpman2}. Since then
there have been numerous studies devoted to the corresponding governing equations,
which are known as the Karpman equations.  For example, the latter admit shock-like 
solutions as well as envelope light solitons \cite{Karpman1,Karpman2,Karpman3}. 
Recently, motivated by observational and experimental studies 
\cite{Ergun-etal,Singh-Lakhina,Kadomtsev-Pogutse,Jones-etal,Choe-etal}, 
the governing equations for light interacting with electron-acoustic waves (EAWs) 
were derived \cite{Shukla-etal}, using a two-species electron fluid model. Since ions
are assumed immobile in Shukla {\it et al.} \cite{Shukla-etal}, the phenomena of 
light-EAW interactions takes place on a timescale shorter than the ion plasma period. 

In this Brief Communication, we will develop a formalism suited for treating effects
of partial coherence of the light source. The modulational instability properties 
of the resulting system of equations are analyzed, and the coherent and incoherent 
cases are contrasted. It is shown that partial coherence lead to damping of
the modulational instability. However, in the limit of short wavelengths, the effect
of a finite spectral width is suppressed, and such short wavelength modes
are therefore more likely to yield, e.g.\ pulse filamentation. The results can
be of importance in space plasmas and laser-plasma systems involving short laser pulses.

In Ref.\ \cite{Shukla-etal}, the equations governing the nonlinear interaction
between coherent light waves and electron-acoustic waves were obtained. 
We consider the propagation of electromagnetic waves in a two population
electron plasma, where one of the electron species is hot, while the other 
electron species is cold. The electromagnetic wave is given by 
the vector potential $A(t,z) = \psi(t,z)\exp(ik_0z - i\omega_0t)$. The evolution of
the slowly varying light wave envelope $\psi(t,z)$ is given by the equation
\begin{equation}\label{eq:nlse}
  2i\omega\left( \frac{\partial}{\partial t} - v_g\frac{\partial}{\partial z} \right)\psi
  + c^2\frac{\partial^2\psi}{\partial z^2} - \omega_{ph}^2(N_h + N_c\delta)\psi = 0 ,
\end{equation}
where $v_g = k_0c^2/\omega_0$ is the group velocity of the light wave, 
$N_h = n_h/n_{h0}$, $N_c = n_c/n_{c0}$, 
$\delta = n_{c0}/n_{h0}$, $n_c$ ($n_h$) is the cold (hot) electron number 
density perturbation, $n_{c0}$ ($n_{h0}$) is the cold (hot) electron background number 
density, and $\omega_{ph} = (4\pi n_{h0}e^2/m_e)^{1/2}$ is the hot electron
plasma frequency. Moreover, the hot electron density distribution is given by
the Boltzmann distribution, i.e.
\begin{equation}\label{eq:hot}
  N_h = \exp(\varphi - \Psi) -1 ,
\end{equation}
where $\varphi = e\phi/T_h$ is the normalized electron-acoustic wave potential 
and $\Psi = e^2|\psi|^2/2m_ec^2T_h$ is the normalized ponderomotive 
force potential due to the light source. The cold electrons are determined by
\begin{subequations}
\begin{equation}\label{eq:cold}
  \frac{\partial N_c}{\partial t} + \frac{\partial}{\partial z}[(1 + N_c)v_c] = 0,
\end{equation}
and
\begin{equation}
  \left(\frac{\partial}{\partial t} + v_c\frac{\partial}{\partial z} \right)v_c 
  = V_{Th}^2\frac{\partial(\varphi - \Psi)}{\partial z} ,
\end{equation}
\label{eq:fluid}
\end{subequations}
where $V_{Th} = (T_h/m_e)^{1/2}$ is the thermal speed of the hot electrons, and 
$v_c$ denotes the cold electron fluid velocity. 
Finally, the electron-acoustic potential is determined by Poisson's equation
\begin{equation}\label{eq:poisson}
  \lambda_{Dh}^2\frac{\partial^2\varphi}{\partial z^2} = N_h + N_c\delta ,
\end{equation}
where $\lambda_{Dh} = (T_h/4\pi n_{h0}e^2)^{1/2}$ is the hot electron 
Debye length. 

From Eqs.\ (\ref{eq:hot})--(\ref{eq:poisson}) we obtain, using 
$|\varphi - \Psi| \ll 1$, the equation \cite{Shukla-etal}
\begin{equation}\label{eq:potential}
  \left( \frac{\partial^2}{\partial t^2} - C_e^2\frac{\partial^2}{\partial z^2}
    - \lambda_{Dh}^2\frac{\partial^4}{\partial t^2\partial z^2} 
  \right)\varphi 
  = \left( \frac{\partial^2}{\partial t^2} - C_e^2\frac{\partial^2}{\partial z^2} \right)\Psi ,
\end{equation}
where $C_e = V_{Th}\delta^{1/2}$. 
Thus, Eqs.\ (\ref{eq:nlse}), (\ref{eq:poisson}) and (\ref{eq:potential}) form
the desired system of equations for coherent interaction between 
light and EAWs. In Ref.\ \cite{Shukla-etal} it has been shown that 
(\ref{eq:nlse}), (\ref{eq:poisson}), and (\ref{eq:potential}) admits
localization and collapse of light pulses. 

In order to account for the effects of partial coherence in the light source, we employ 
the Wigner formalism. Starting from the two-point correlation function of the field
of interest (in our case the light field), one performs a Fourier transform of this
correlation function and obtains a generalized distribution function \cite{Wigner}. 
Thus, we let the generalized distribution function for the light quanta be 
\cite{Wigner}
\begin{equation}\label{eq:transf}
  \rho(t,z,k) = \frac{1}{2\pi}\int\,d\zeta\,e^{ik\zeta}%
    \langle \psi^*(t, z + \zeta/2)\psi(t, z - \zeta/2) \rangle ,
\end{equation}
where the angular brackets denotes the ensemble average. Then the
light intensity is given by 
\begin{equation}\label{eq:intensity}
  \Psi = \frac{e^2|\psi|^2}{2m_ec^2T_h} = \frac{e^2}{2m_ec^2T_h}\int\,dk\,\rho(t,z,k) .
\end{equation}
Applying the time derivative on the definition (\ref{eq:transf}),
and using Eq.\ (\ref{eq:nlse}), we obtain the kinetic equation
\begin{equation}\label{eq:kinetic}
  \frac{\partial\rho}{\partial t} + \left(v_g + \frac{c^2k}{\omega_0}\right)\frac{\partial\rho}{\partial z}
  - \frac{\omega_{ph}^2\lambda_{Dh}^2}{\omega_0}\left( \frac{\partial^2\varphi}{\partial z^2}
   \right)%
    \sin\left( \frac{1}{2}\stackrel{\leftarrow}{\frac{\partial}{\partial z}}
    \stackrel{\rightarrow}{\frac{\partial}{\partial k}} \right)\rho =  0 
\end{equation}
for the light pseudo-particles. Here the arrows denotes the direction 
of operation, and the $\sin$-operator is defined in terms of
its Taylor expansion. By assuming a light distribution function
with a finite spectral width, i.e. a finite spread in the light power spectrum, 
the effect of partial coherence can be
incorporated into the light propagation through the two-electron plasma.
Thus, Eqs.\ (\ref{eq:potential}), (\ref{eq:intensity}),
and (\ref{eq:kinetic}) describes the interaction between partially
coherent light and EAWs. 

Next  we analyze the stability properties of the system (\ref{eq:potential}), 
(\ref{eq:intensity}), and (\ref{eq:kinetic}) by perturbing these equations around a stationary 
plasma state (see, e.g. \cite{Anderson-etal} for
a similar treatment of quantum plasmas). We let 
$\rho(t,z,k) = \rho_0(k) + \rho_1(k)\exp(iKz - i\Omega t)$, where
$|\rho_1| \ll |\rho_0|$, and $\varphi = \varphi_1\exp(iKz - i\Omega t)$. Moreover, we
have $\Psi = \Psi_0 + \Psi_1\exp(iKz - i\Omega z)$, where $|\Psi_1| \ll |\Psi_0|$. 
Linearizing the system (\ref{eq:potential}), (\ref{eq:intensity}), and (\ref{eq:kinetic})
with respect to the first order quantities, we obtain the nonlinear dispersion relation
\begin{equation}\label{eq:dispersion}
  1 = \frac{e^2}{2m_ec^2T_h}\frac{\omega_{ph}^2\lambda_{Dh}^2}{2\omega_0}%
    K^2\Lambda\int\,dk\,\frac{\rho_0(k + K/2) - \rho_0(k - K/2)}%
      {\Omega - (v_g + c^2k/\omega_0)K} ,
\end{equation}
where $\Lambda = (\Omega^2 - C_e^2K^2)/(\Omega^2 - C_e^2K^2 
+ \lambda_{Dh}^2K^2\Omega^2)$. 

In the case of coherent light we have $\rho_0 = |\psi_0|^2\delta(k - \kappa_0)$,
and the nonlinear dispersion relation (\ref{eq:dispersion}) yields
\begin{equation}\label{eq:mono}
  1 = -\frac{e^2}{2m_ec^2T_h}\frac{\omega_{ph}^2\lambda_{Dh}^2}{2c^2}%
    K^2\Lambda \frac{|\psi_0|^2}%
      {[\kappa_0 - \omega_0(\Omega - v_gK)/c^2K]^2 - K^2/4} ,
\end{equation}
i.e.
\begin{equation}\label{eq:mono2}
  (\Omega^2 - C_e^2K^2 + \lambda_{Dh}^2K^2\Omega^2)%
  \left\{
  \left[\Omega - \left(v_g +\frac{c^2\kappa_0}{\omega_0}\right)K \right]^2 
  - \frac{c^4K^4}{4\omega_0^2}  
  \right\}
  = -\Psi_0\frac{c^2\omega_{ph}^2\lambda_{Dh}^2}{2\omega_0^2}%
    K^4(\Omega^2 - C_e^2K^2),
\end{equation}

Suppose now that the pulse phase $\Phi(z)$ experiences a random 
variation, which satisfies \cite{Loudon} 
$\langle \exp[-i\Phi(z + \zeta/2)]\exp[i\Phi(z - \zeta/2)]\rangle 
= \exp(-\Delta |\zeta| )$, where $2\Delta$ is the full wavenumber width at half 
maximum of the power spectrum. Then the corresponding distribution
function is given by the Lorentzian \cite{Loudon}
\begin{equation}\label{eq:lorentz}
  \rho_0(k) = \frac{|\psi_0|^2}{\pi}\frac{\Delta}{(k - \kappa_0)^2 + \Delta^2} ,
\end{equation}
where $\kappa_0$ is a wavenumber shifting the location of the maxima of the 
distribution function $\rho_0{k}$ 
Inserting the expression (\ref{eq:lorentz}) into the nonlinear dispersion
relation (\ref{eq:dispersion}) we obtain
\begin{equation}\label{eq:spectrum}
  1 = -\frac{e^2}{2m_ec^2T_h}\frac{\omega_{ph}^2\lambda_{Dh}^2}{2c^2}%
    K^2\Lambda \frac{|\psi_0|^2}%
      {[\kappa_0 -i\Delta - \omega_0(\Omega - v_gK)/c^2K]^2 - K^2/4} ,
\end{equation}
i.e.
\begin{eqnarray}
  &&
  (\Omega^2 - C_e^2K^2 + \lambda_{Dh}^2K^2\Omega^2)%
  \left\{
  \left[\Omega - \left(v_g +\frac{c^2(\kappa_0 - i\Delta)}{\omega_0}\right)K \right]^2 
  - \frac{c^4K^4}{4\omega_0^2}  
  \right\}
  \nonumber \\ &&\qquad\qquad 
  = -\Psi_0\frac{c^2\omega_{ph}^2\lambda_{Dh}^2}{2\omega_0^2}%
    K^4(\Omega^2 - C_e^2K^2),
\label{eq:spectrum2}
\end{eqnarray}
Comparing the dispersion relations (\ref{eq:mono2}) and (\ref{eq:spectrum2}),
it can be seen that the spectral broadening will reduce the growth rate.  

\begin{figure}[ht]
  \includegraphics[width=0.8\textwidth]{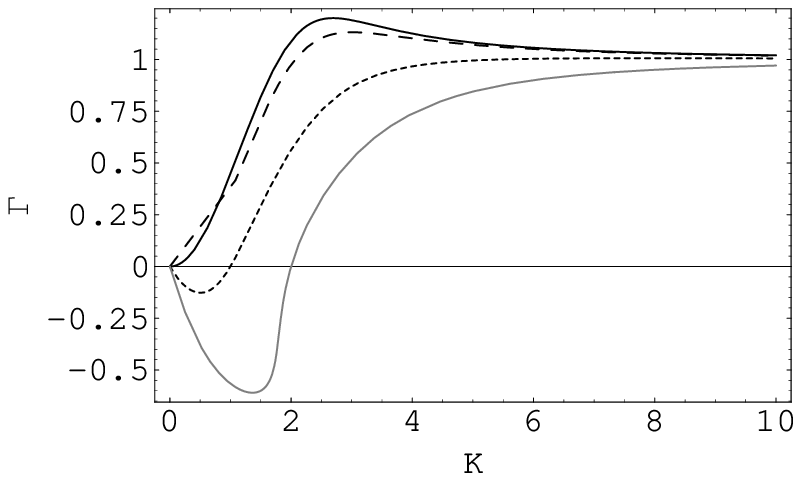}
  \caption{The growth rate $\Gamma$ plotted as a function of the wavenumber
  $K$ with the wavenumber shift $\kappa_0 = 0$. 
  Here we have used the normalized variables (\ref{eq:norm}). Moreover,
  we have used the values normalized values $C_e = 0.5$, $\lambda_{Dh} = 0.5$,
  $\Psi_0 = 0.5$. The solid black curve has $\Delta = 0$, the dashed curve $\Delta = 0.1$,
  the dotted curve $\Delta = 0.5$, and the solid gray curve uses $\Delta = 1$. The damping
  effect of a finite spectral width can clearly be seen, although the short wavelength
  modes are less affected. The asymptotic value of the growth rate in this case is $1$.}
  \label{fig:1}
\end{figure}

From now on we put the wavenumber shift $\kappa_0$ to zero, thus
centering the Lorentzian distribution (\ref{eq:lorentz}) around $0$, as well as 
transforming to a comoving system, such that $v_g \rightarrow 0$. 
We may use normalized and dimensionless variables defined by
\begin{equation} \label{eq:norm}
  K \rightarrow \frac{cK}{\omega_0} , \quad
  \Omega \rightarrow \frac{\Omega}{\omega_0} , \quad
  C_e \rightarrow \frac{C_e}{c} , \quad
  \lambda_{Dh} \rightarrow \frac{\omega_0\lambda_{Dh}}{c} , \quad
  \Psi_0 \rightarrow \frac{\omega_{ph}^2\lambda_{Dh}^2\Psi_0}{2c^2} ,
\text{ and } \Delta \rightarrow \frac{c\Delta}{\omega_0}.
\end{equation}
Using these dimensionless variables, we have plotted the typical behavior of
the growth rate $\Gamma = i(\mathrm{Re}\,\Omega - \Omega)$, as given by 
the dispersion relations (\ref{eq:mono2}) and (\ref{eq:spectrum2}),
in Fig.\ \ref{fig:1}. The uppermost
curve corresponds to $\Delta = 0$, i.e.\ the dispersion relation as given by 
(\ref{eq:mono}). The growth rate asymptotically tends to $1$ for the
parameter values chosen in Fig.\ \ref{fig:1}, and characteristically 
approaches an constant value for large wavenumbers if other parameter
values are chosen. Moreover, 
the three lower curves has successively higher values of
the spectral width $\Delta$. The damping character of a nonzero spectral 
width can clearly be seen. Furthermore, it is clear that for long wavelengths,
i.e.\ small values of $K$, the damping may completely suppress the growth
rate, although small values may even enhance the growth rate (cf.\ the solid and dashed 
curved for small $K$). However, for short wavelength modes, i.e.\ higher values of $K$, 
even the cases with nonzero $\Delta$ asymptotically tends to the same 
value of the growth rate as in the monochromatic case. Thus, short wavelength
modes seem almost unaffected by the spectral broadening, and such short
wavelength perturbations would therefore dominate in the contribution to the 
filamentation, as well as 
in soliton and shock wave formation \cite{Shukla-etal} of the incoherent 
light interacting with EAWs. In principle, these observations could be of importance in
situations where the two-species electron model plays an important role, such as in 
space plasmas and laser-plasma systems 
\cite{Ergun-etal,Singh-Lakhina,Kadomtsev-Pogutse,Jones-etal,Choe-etal}.

To summarize, we have derived the governing equations for the interaction between
light (coherent and incoherent) and EAWs. The limit of mono-chromatic light was analyzed, 
and compared to the case where spectral broadening was taken into account.
It was shown that the effect of a finite width in the power spectrum of the light 
in general was to suppress 
the modulational instability. However, for short wavelength modes the
spectral broadening was shown only to have a small influence on the 
modulational instability growth rate, thus making these modes more
likely to dominate in the filamentation of pulses and the formation
of shocks and solitons in space and laser plasmas. 

\newpage



\newpage

\end{document}